\documentclass[aps,floatfix,twocolumn,showpacs,showkeys,preprintnumbers,amsmath]{revtex4}

\usepackage{dcolumn}
\usepackage{bm}
\usepackage{graphicx}
\usepackage{color}
\graphicspath{{Images/}}

\newcommand{\la}{\langle}
\newcommand{\ra}{\rangle}

\begin{document}

\title{Role of correlations on spin-polarized neutron matter}
\author{ Isaac Vida\~na$^1$, Artur Polls$^{2}$ and Victoria Durant$^{3,4}$}
\affiliation{$^1$CFisUC, Department of Physics, University of Coimbra, PT-3004-516
Coimbra, Portugal}
\affiliation{$^2$Departament d'Estructura i Constituents de la Mat\`eria and
Institut de Ci\`encies del Cosmos,
Universitat de Barcelona, Avda. Diagonal 647, E-08028 Barcelona, Spain}
\affiliation{$^3$ Institut f\"{u}r Kernphysik. Technische Universit\"{a}t Darmstadt, 64289 Darmstadt, Germany}
\affiliation{$^4$ ExtreMe Matter Institute EMMI, GSI Helmholtzzentrum f\"{u}r Scherwionenforschung GmbH, 64291, Germany}

\newcommand{\m}{\multicolumn}
\renewcommand{\arraystretch}{1.2}

\begin{abstract}
\begin{description}
\item[Background:]
The possible existence of a phase transition to a ferromagnetic state in neutron matter as origin of the extremely high
magnetic fields of neutron stars is still an open issue. Whereas many phenomenological interactions predict this transition
at densities accesible in neutron stars, microscopic calculations based on realistic interactions show no indication of it.  
The existence or non-existence of this transition is a consequence of the different role of nucleon-nucleon correlations in 
polarized and unpolarized
neutron matter. Therefore, to give a definite answer to this issue it is necessary to analyze the behavior of these correlations.

\item[Purpose:]
Using the Hellmann--Feynman theorem we analyze the contribution of the different terms of the nucleon-nucleon
interaction to the spin symmetry energy of neutron matter with the purpose of identifying the nature and role of correlations
in polarized and unpolarized neutron matter.
\item[Methods:]
The analysis is performed  
within the microscopic Brueckner--Hartree--Fock approach using the Argonne V18 realistic potential plus the 
Urbana IX three-body force.
\item[Results:] Our results show no indication of a ferromagnetic transition as the spin symmetry energy of neutron matter is always an increasing function of density. 
They show also that the main contribution to it 
comes from the S=0 channel, acting only in non-polarized neutron matter, in particular from the $^1S_0$ and the $^1D_2$ partial waves. Three-body forces 
are found to play a secondary role in the determination of the spin symmetry energy.
\item[Conclusions:] By evaluating the kinetic energy difference between the correlated system and the underlying Fermi sea to estimate the importance of correlations 
in spin-polarized neutron matter, we conclude that non-polarized neutron matter is more correlated than totally polarized one.

\end{description}
\end{abstract}


\maketitle  


\section{Introduction}


The spin symmetry energy of neutron matter, defined as the difference between 
the energy per particle of spin polarized and unpolarized neutron matter, 
is the main ingredient to understand the spin susceptibility of
 neutron matter, which is basically  proportional to the invers of this quantity. Microscopic  calculations of the spin susceptibility, using realistic 
interactions and a variety of many-body methods show that the correlations
 induced by these realistic interactions considerable reduce the spin-susceptibility with respect
 to the underlying non-interacting Fermi seas \cite{qmc,vidana02,vidana02b,vidana06,dbhf,dbhf2,locv}. This reduction implies an increase of the spin-symmetry energy of neutron matter. 
This prediction has also important consequences in  the description of situations of
 astrophysical interest, such as for instance, the calculation of the mean free path of
 neutrinos in dense matter and, in general, the study of supernovae and protoneutron stars \cite{reddy1999}. 

In contrast with this scenario, it has been theoretically speculated that the spin symmetry of neutron matter can become zero, a fact that would indicate the existence of a phase transition to a 
ferromagnetic state \cite{BROWNELL,RICE,CLARK,CLARK2,SILVERSTEIN,OST,PEAR,PANDA,BACK,HAENSEL,JACK,KUT,MARCOS}.  Notice, that looking at the kinetic energies of the corresponding
 underlying Fermi seas,  at a given density the kinetic energy of the polarized Fermi sea will always be  larger than the unpolarized one. Therefore, the hypothetical ferromagnetic transition should be
 a consequence of the different role of the interactions in polarized and unpolarized 
neutron matter. In fact, many effective nuclear interactions of Skyrme \cite{VIDAURRE,rios2005} or Gogny \cite{lopez-val2006} type predict
 this  transition at densities accesible in neutron stars.
However, in accordance with the reduction of the spin susceptibility comented above, microscopic calculations based on realistic interactions do not predict such transition 
at least in the wide range of densities which have been explored \cite{qmc,vidana02,vidana02b,vidana06,dbhf,dbhf2,locv}. 
The study of spin-polarization has also been recently considered for nuclear matter and finite nuclei 
using finite range effective interactions \cite{vinas}. The possibility of a ferromagnetic 
transition has also been discussed in the context of  hard-sphere systems  in conexion with ultracold atom systems \cite{pilati2010,arias2012}. 
All these facts have motivated the interest for the study of  neutron matter and in particular of polarized neutron matter. 

It has also been  pointed out that due to the large value of the $^1S_0$ scattering length, 
the behaviour  of neutron matter, at densities where the physics is dominated by this partial wave, 
should show some similarities with the behaviour of a unitary Fermi gas \cite{maieron2008}. At the same time, the 
absence due to the Pauli principle  of the $^1$S$_0$ channel  in polarized neutron matter has driven 
the question of up to which density  polarized neutron matter can behave as a weakly interacting Fermi gas \cite{kruger2015}.

Motivated by these questions, we have performed a microscopic calculation, in the framework
 of the 
Brueckner--Hartree--Fock approximation, of the magnetic susceptibility of neutron matter
 employing  the Argonne V18 (Av18)  realistic nucleon-nucleon interaction \cite{WIRINGA} 
suplemented with 
 the Urbana IX three-body force \cite{PUDLINER}, which for the use in the BHF approach is reduced to an effective two-body density-dependent interaction by averaging over the third nucleon \cite{bf99}. 
In order  to identify the nature of the correlations  responsible 
for the behavior of the magnetic susceptibility we have analyzed the contributions to the spin symmetry energy of neutron matter  of
 the different partial waves and also of the different operatorial parts of the interaction. 
In addition, the degree of correlation of the two systems, polarized and unpolarized
 neutron matter, is discussed by comparing the differences of the kinetic energy of the correlated system  with the ones of the underlying Fermi sea.  

The paper is organized in the following way. In Sec. II the Brueckner--Hartree--Fock approach
to spin polarized neutron matter  and the Hellmann-Feynman theorem \cite{hellmann,feynman}
are shortly reviewed. Results for the 
magnetic susceptibility or, equivalently, for the spin symmetry energy and its density dependence
 are presented in Sec. III, where is also discussed the contribution of the different partial waves. 
Finally, a short summary and the main conclusions are given in Sec. IV. 


\section{BHF approach of spin-polarized neutron matter}

Spin-polarized neutron matter is an infinite nuclear system made of two different fermionic components: neutrons with
spin up and neutrons with spin down, having densities $\rho_\uparrow$ and $\rho_\downarrow$, respectively. The total density of the system is given by
\begin{equation}
\rho=\rho_\uparrow+\rho_\downarrow \ .
\label{eq:den}
\end{equation}
The degree of spin polarization of the system can be expressed by means of the spin polarization $\Delta$ defined as
\begin{equation}
\Delta=\frac{\rho_\uparrow-\rho_\downarrow}{\rho} \ .
\label{eq:sp}
\end{equation}
Note that the value $\Delta=0$ corresponds to nonpolarized (NP) or paramagnetic ($\rho_\uparrow=\rho_\downarrow$) neutron matter, whereas 
$\Delta=\pm 1$ means that the system is totally polarized (TP), {\it i.e.,} all the spins are aligned along the same direction. Partially 
polarized states correspond to values of $\Delta$ between $-1$ and $+1$.

The energy per particle of spin-polarized neutron matter does not change when a global flip of the spins is performed. Therefore, it can be 
expanded on the spin polatization $\Delta$ as
\begin{equation}  
E(\rho,\Delta) = E_{NP}(\rho)+S_{sym}(\rho)\Delta^2 + S_{4}(\rho)\Delta^4 + {\cal O}(6) \ ,
	\label{eq:en}
	\end{equation}
	where $E_{NP}(\rho)\equiv E(\rho,0)$ is the energy per particle of nonpolarized neutron matter, $S_{sym}(\rho)$
	is defined as the spin symmetry energy, 
	\begin{equation}  
	S_{sym}(\rho)=\frac{1}{2}\frac{\partial^2 E(\rho,\Delta)}{\partial \Delta^2}\Big|_{\Delta=0}
	\label{eq:ssym}
	\end{equation}
	and
	\begin{equation}  
	S_{4}(\rho)=\frac{1}{24}\frac{\partial^4 E(\rho,\Delta)}{\partial \Delta^4}\Big|_{\Delta=0} \ .
	\label{eq:s4}
	\end{equation}
	It has been shown (see {\it e.g.,} Refs. \cite{vidana02,vidana02b,vidana06}) that the energy per particle of spin-polarized neutron matter 
	is practically parabolic in the full range of spin polarizations. Therefore, contributions from $S_4(\rho)$ and other higher order
	terms can be neglected, and one can, in good approximation, estimate the spin symmetry energy simply as 
	the difference between the energy per particle of totally polarized, $E_{TP}(\rho) \equiv E(\rho,\pm 1)$,  and nonpolarized 
	neutron matter {\it i.e.,}
	\begin{equation}
	S_{sym}(\rho) \sim E_{TP}(\rho)-E_{NP}(\rho) \ .
	\label{eq:ssym2}
	\end{equation}

	A particularly interesting macroscopic property of spin polarized neutron matter related to $S_{sym}(\rho)$ is the magnetic susceptibility  $\chi(\rho)$
	which, at each density, characterizes the response of the system to an external magnetic field and gives a measure of the energy 
	required to produce a net spin alignment in the direction of it. If the strengh of the field is small $\chi(\rho)$ can be obtained simply as (see {\it e.g.,} Ref.\ \cite{vidana02})
	\begin{eqnarray}
	\chi(\rho)&=&\frac{\mu^2 \rho}{\frac{\partial^2 E(\rho,\Delta)}{\partial \Delta^2}\Big|_{\Delta=0}} \nonumber \\
		    &=&\frac{\mu^2 \rho}{2S_{sym}(\rho)}
		    \label{eq:chi1}
		    \end{eqnarray}
		    where $\mu$ is the magnetic moment of the neutron and in the second equality we have used Eq.\ (\ref{eq:ssym}).

		    The BHF description of spin-polarized neutron matter starts with the construction of the neutron-neutron $G$-matrix, which
		    describes in an effective way the interaction between two neutrons for each one of the spin combinations $\uparrow\uparrow,
		    \uparrow\downarrow, \downarrow\uparrow$ and $\downarrow\downarrow$. This is formally obtained by solving the well-known 
		    Bethe-Goldstone equation, written schematically as
		    \begin{eqnarray}
		    G(\omega)_{\sigma_1\sigma_2\sigma_3\sigma_4}&=&
		    V_{\sigma_1\sigma_2\sigma_3\sigma_4} 
		    +\frac{1}{\Omega}\sum_{\sigma_i\sigma_j}
		    V_{\sigma_1\sigma_2\sigma_i\sigma_j} \nonumber \\
			    &\times&\frac{Q_{\sigma_i\sigma_j}}{\omega-\varepsilon_{\sigma_i}-\varepsilon_{\sigma_j}+i\eta}
			    G(\omega)_{\sigma_i\sigma_j\sigma_3\sigma_4} \ ,
			    \label{eq:bge}
			    \end{eqnarray}
			    where $\sigma=\uparrow, \downarrow$ indicates the spin projection of the two neutrons in the 
			    initial, intermediate and final states, $V$ is the bare nucleon-nucleon interaction, $\Omega$ is the (large) volume 
			    enclosing the system, $Q_{\sigma_i\sigma_j}$ is the Pauli operator taking into account the effect of the 
			    exclusion principle on the scattered neutrons, and $\omega$ is the so-called starting energy defined
			    as the sum of the non-relativistic single-particles energies, $\epsilon_{\uparrow(\downarrow)}$, of the interacting neutrons.
			    We note that Eq.\ (\ref{eq:bge}) is a coupled channel equation.

			    The single-particle energy of a neutron with momentum ${\vec k}$ and spin projection $\sigma$ is given by
			    \begin{equation}
			    \epsilon_{\sigma}({\vec k})=\frac{\hbar^2
				    k^2}{2m}+\mbox{Re}[U_{\sigma}({\vec k})] \ ,
				    \label{spe}
				    \end{equation}
				    where the real part of the single-particle potential
				    $U_{\sigma}({\vec k})$  represents the average potential ``felt'' by
				    a neutron due to its interaction with the other neutrons of the system.  In the BHF approximation
				    $U_{\sigma}({\vec k})$ is calculated through the ``on-shell'' $G$-matrix, and is given by
				    \begin{equation}
				    U_{\sigma}({\vec k})=\frac{1}{\Omega}
				    \sum_{\sigma'\vec k'}
				    \langle {\vec k}\sigma{\vec k'}\sigma' |G(\epsilon_{\sigma}(\vec k)+\epsilon_{\sigma'}(\vec k'))|{\vec k}\sigma{\vec k'}\sigma'\rangle_{A} \ ,
				    \label{eq:spp}
				    \end{equation}
				    where the sum runs over all neutron up and neutron down occupied states and the matrix elements 
				    are properly antisymmetrized. Once  a self-consistent solution of
				    Eqs.\ (\ref{eq:bge})-(\ref{eq:spp}) is achieved, the energy per particle in the BHF approximation can be calculated as
				    \begin{eqnarray}
				    E_{BHF}(\rho,\Delta)&=&\frac{1}{A}
				    \sum_{\sigma}\sum_{|\vec k|\leq k_{F_{\sigma}}}
				    \frac{\hbar^2
					    k^2}{2m} \nonumber \\
						    &+&
						    \frac{1}{2A}\sum_{\sigma}\sum_{|\vec k|\leq k_{F_{\sigma}}}
						    \mbox{Re}[U_{\sigma}(\vec k)]\ ,
						    \label{eq:bhf}
						    \end{eqnarray}
						    where the first term of the r.h.s. is simply the contribution of the free Fermi gas (FFG), and the second one is sometimes called in the literature {\it correlation energy}. We note that $E_{BHF}$ represents only the sum of {\it two-hole-line} diagrams and includes only the effect of two-body correlations
						    through the $G$-matrix.  It has been shown by Song {\it et al.,} \cite{SONG} that the contribution to the energy from {\it three-hole-line} diagrams (which account for the effect of three-body correlations) is minimized when the so-called continuous prescription \cite{JEKEUNE}
						    is adopted for the single-particle potential when solving the Bethe--Goldstone equation. This presumably enhances the convergence of the hole-line expansion of which the BHF approximation represents the lowest order. We adopt this prescription in our BHF calculations which are
						    done using the Argonne V18 (Av18) potential \cite{WIRINGA} supplemented with
						    the Urbana IX three-nucleon force \cite{PUDLINER}, which for the
						    use in the BHF approach is reduced first to an effective
						    two-nucleon density-dependent force by averaging over the
						    coordinates of the third nucleon \cite{bf99,TBF}.

						    The BHF approach does not give direct access to the separate contributions of the kinetic and potential energies because it does not provide the correlated many-body wave funtion $|\Psi\rangle$. However, it has been shown \cite{muether99, sartor00, sartor01, vidana05} that the Hellmann--Feynman theorem \cite{hellmann,feynman} can be used to estimate the ground-state expectation value of both contributions from the derivative of the total energy with respect to a properly introduced parameter. Writing the nuclear matter Hamiltonian as $H=T+V$, and defining a $\lambda$-dependent Hamiltonian $H(\lambda)=T+\lambda V$, the expectation value of the potential energy is given as
						    \begin{equation}
						    \langle V \rangle \equiv \frac{\langle \Psi | V | \Psi \rangle}{\langle \Psi | \Psi \rangle}
						    =\left(\frac{dE}{d\lambda}\right)_{\lambda=1} 
						    \end{equation}
						    and the kinetic energy contribution $\langle T \rangle$ can be simply obtained by substracting $\langle V \rangle$ from $E_{BHF}$.


						    \section{Results and Discussion}

						    \begin{figure}[t]
						    \begin{center}
						    \includegraphics[width=.45\textwidth]{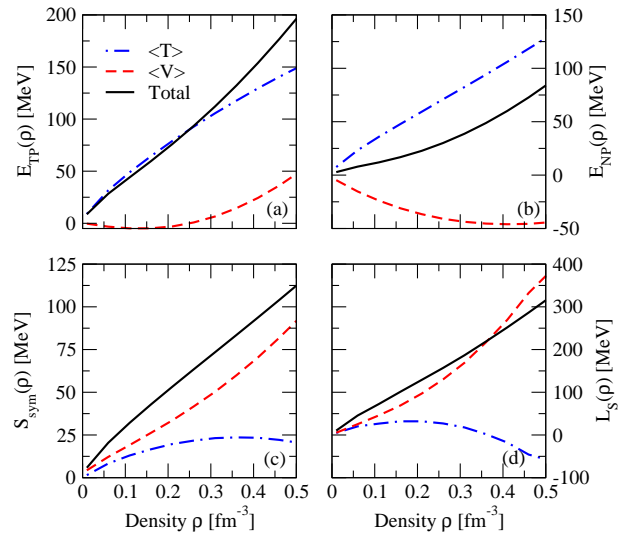}
						    \caption{(color on-line) Kinetic $\la T\ra$ and potential $\la V \ra$ energy contributions to the total energy per particle of totally polarized and nonpolarized
							    neutron matter (panels a and b), and to the spin symmetry energy and its slope parameter (panels c and d) as a function of density. Notice the different scale in the ordinates.} 
							    \label{fig:contri2}
							    \end{center}
							    \end{figure}


							    The discussion of our results starts by showing in Fig.\ \ref{fig:contri2} the density dependence of the kinetic $\langle T \rangle$ and potential $\langle V \rangle$ energy contributions to the 
							    energy per particle of both TP (panel (a)) and NP (panel (b)) neutron matter as well as to the spin symmetry energy (panel (c)) and its slope 
							    parameter (panel (d)) defined as 
							    \begin{equation}
							    L_S(\rho) = 3\rho\frac{\partial S_{sym}(\rho)}{\partial\rho} \, , 
							    \label{ls}
							    \end{equation}
							    in analogy with the slope parameter of the nuclear symmetry energy, $L(\rho)$. The particular values of these contributions at 
							    the empirical saturation density of symmetric nuclear matter, $\rho_0=0.16$ fm$^{-3}$, are reported in Tab.\ \ref{tab1}. 
							    The results have been obtained by applying the Hellmann--Feynman theorem as explained at the end of the previous section.
							    As it can be seen in the figure, the total energy of TP neutron is always more repulsive than the NP one in all the density range explored. This 
							    additional repulsion of TP neutron matter can be understood, firstly, in terms of the kinetic energy contribution, which is
							    larger in the TP case than in the NP one. Secondly, in terms of the potential
							    energy one because, due to symmetry arguments, all partial waves with even orbital angular momentum $L$ (some of them
									    attractive, as the important $^1S_0$) are excluded in TP neutron matter (see Tab.\ \ref{tab2}). An interesting conclusion which can
							    be inferred from here, already pointed out in previous works of the authors \cite{vidana02,vidana02b,vidana06} and other 
							    studies \cite{qmc,dbhf,dbhf2,locv}, is that a spontaneous phase transiton to a ferromagnetic state is not to be expected. If such 
							    a transition would exist, a crossing of energies of the TP and NP systems, with the consequent change of the sign of the spin symmetry energy,
							    would be observed at some density, indicating that the ground state of the system would be ferromagnetic from that density on. Notice that there is 
							    no sign of such a crossing on the figure and that, on the contrary, it becomes less and less favorable as the density increases.
							    As it is seen in the figure the kinetic enegy contribution to the spin symmetry energy, although it is always smaller than that of the potential energy one, is  
							    not negligible and, in particular, amounts $\sim 38\%$ of its total value at $\rho_0$. This result is different from what is found in the case of the nuclear symmetry 
							    energy, $E_{sym}(\rho)$. In this case the kinetic energy contribution to $E_{sym}(\rho)$ is very small (and even negative) due to the strong cancellation of the kinetic energies
							    of neutron and symmetric nuclear matter \cite{providencia,carbone1}. Finally, note that the slope parameter $L_S(\rho)$ is also clearly dominated in the whole density range by the potential 
							    energy contribution ($\sim 75 \%$ at $\rho_0$) except at very low densities where the kinetic energy one is of similar order. 
							    Also interesting is the fact that in a significative density region around $\rho_o$, $L_s(\rho)$ is rather linear, indicating that 
the derivative of $S_{sym}(\rho)$ respect to the density is  approximately constant (see Eq.\ (\ref{ls})).

							    \begin{center}
							    \begin{table}[t]
							    \begin{tabular}{crrrr}
							    \hline
							    \hline
							    & $E_{TP}$ & $E_{NP}$ & $S_{sym}$ & $L_S$ \\
								    \hline
								    $\langle T_{FS} \rangle$                  &  $55.669$ &  $35.069$ & $20.600$ & $ 41.200$ \\
									    $\langle T \rangle$                       &  $64.452$ &  $47.827$ & $16.625$ & $25.225$ \\
									    $\langle V \rangle$                       &  $-4.784$ &  $-31.050$ & $26.266$ & $75.914$ \\
									    Total                                     &  $59.668$ &  $16.777$  & $42.891$ & $101.139$ \\
									    \hline
									    \hline
									    \end{tabular}
									    \caption{Kinetic, $\langle T \rangle$, and potential, $\langle V \rangle$, contributions
										    to the total energy per 
											    particle of totally polarized (TP) and non-polarized (NP) neutron matter at the
											    empirical salturation density of symmetric nuclear matter, $\rho_0=0.16$ fm$^{-3}$. 
											    The contribution to the corresponding spin symmetry energy $S_{sym}$ and its 
											    slope parameter $L_S$ are reported in the last two columns, respectively. $<T_{FS}>$ correspond to the results
of the unerlying Fermi seas. Results are given in MeV. }
											    \label{tab1}
											    \end{table}
											    \end{center}
											    To get a further physical insight on the role of the potential energy, it is useful to look at the spin channel and partial wave decomposition of its contribution to the energies of TP and NP neutron matter as well as that to 
											    the spin symmetry energy and its slope parameter. These contributions are denoted as $\langle V \rangle_{TP}$,  $\langle V \rangle_{NP}$, $S^{\langle V \rangle}_{sym}$ and  $L^{\langle V \rangle}_S$, respectively, and their values at $\rho_0$ are shown in Tab. \ref{tab2}. The main contribution to 
											    $S^{\langle V \rangle}_{sym}$ and $L^{\langle V \rangle}_S$ is that of the $S=0$ channel, acting only in NP neutron matter, and in particular that of the $^1S_0$ and $^1D_2$ partial waves, which at $\rho_0$ amount $\sim 99\%$ of $S^{\langle V \rangle}_{sym}$ and $\sim 70\%$ of 
											    $L^{\langle V \rangle}_S$. Notice that, at this density, the contribution of the $S=1$ channel to the energies of TP and NP matter is very similar and, therefore, the 
											    contribution of this channel to  $S^{\langle V \rangle}_{sym}$ is almost negligible. This is mainly due to the strong compensation of the P- and the F-waves which almost cancel completely, and to
											    the small contribution of the H- and J- and L-waves. Note also that, for this reason, the contribution  from those partial waves where the tensor force is active ($^3P_2, ^3F_2, ^3F_4, ^3H_4, ^3H_6, ^3J_6, ^3J_8, ^3L_8$) represents a small percentage of the total values of $S^{\langle V \rangle}_{sym}$ and $L^{\langle V \rangle}_S$. This can interpreted as an indication that the tensor force plays a minor role in  the determination of the spin symmetry energy  and its density dependence. This conclusion differs from that drawn in the case of the nuclear symmetry energy whose value at saturation and its density dependence is known to be clearly dominated by the tensor force \cite{carbone,vidana2009} (see also {\it e.g.,} Ref.\ \cite{epja} and references therein).
											    \begin{center}
											    \begin{table}[t]
											    \begin{tabular}{crrrr}
											    \hline
											    \hline
											    & $\langle V \rangle_{TP}$ & $\langle V \rangle_{NP}$ & $S^{\langle V \rangle}_{sym}$ & $L^{\langle V \rangle}_S$ \\
												    \hline
												    $S=0$           &  $0$ &  $-26.875$ &  $26.875$ & $56.198$ \\
												       $S=1$           &  $-4.784$ &  $-4.175$ & $-0.609$ & $19.716 $ \\
												       \hline
												       $^1S_0$   &  $0.000$        & $-21.432$  & $21.432$   & $32.086$     \\

												       $^3P_0$   &  $-5.499$      & $-4.624$   &  $-0.875$   & $3.313$    \\
													       $^3P_1$   &  $19.644$      & $13.027$   & $6.617$     & $30.927$    \\
													       $^3P_2$   &  $-19.915$    & $-13.299$ & $-6.616$   & $-14.966$   \\
													       $^1D_2$   &  $0.000$       & $-4.787$   & $4.787$   & $21.185$    \\

													       $^3F_2$   &  $-1.263$      & $-0.574$   & $-0.689$   & $-2.655$   \\
														       $^3F_3$   &  $3.109$        & $1.639$    &  $1.470$    & $5.253$   \\
														       $^3F_4$   &  $-1.726$      & $-0.597$  &  $-1.129$   & $-5.492$   \\
														       $^1G_4$   &  $0.000$       & $-0.607$   & $0.607$   & $3.055$    \\

														       $^3H_4$   &  $-0.042$     & $0.012$     & $-0.054$  & $-0.094$   \\
															       $^3H_5$   &  $0.699$       & $0.186$     & $0.513$    & $1.889$   \\
															       $^3H_6$   &  $-0.028$     & $0.024$     & $-0.052$    & $-0.345$   \\
															       $^1I_6$    &  $0.000$       & $-0.059$    & $0.059$    & $0.116$    \\

															       $^3J_6$    &  $0.051$       & $0.024$     & $0.027$    & $0.370$   \\
																       $^3J_7$    &  $0.107$       & $-0.025$    & $0.132$   & $0.476$   \\
																       $^3J_8$    &  $0.050$       & $0.020$     & $0.030$    & $0.332$   \\
																       $^1K_8$   &  $0.000$       & $0.011$     & $-0.011$    & $0.245$    \\

																       $^3L_8$   &  $0.029$       & $0.011$     & $0.018$    & $0.219$   \\
																	       \hline
																	       \hline

																	       \end{tabular}
																	       \caption{Spin channel and partial wave decomposition of the potential energy of TP and NP neutron matter at $\rho_0=0.16$ fm$^{-3}$. The decompostions of the potential energy contribution to the spin symmetry and its slope parameter are also shown. Results are given in MeV.}
																	       \label{tab2}
																	       \end{table}
																	       \end{center}
																	       \begin{figure}[t]
																	       \begin{center}
																	       \includegraphics[width=.45\textwidth]{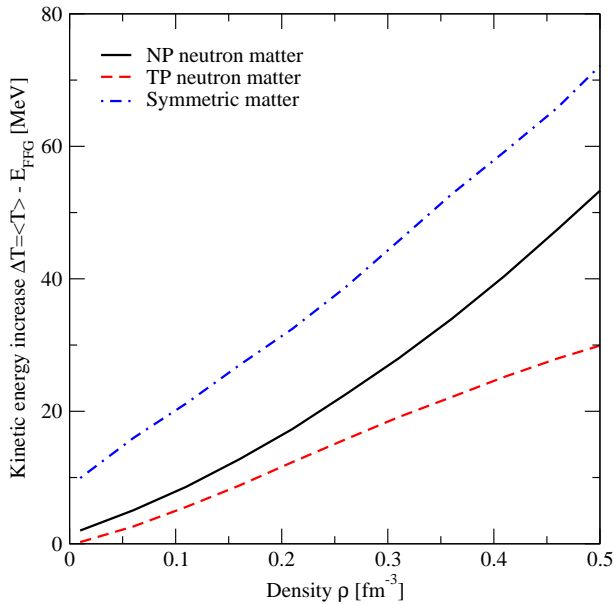}
																	       \caption{(color on-line) Increase of the kinetic energy per particle due to SRC in TP and NP neutron matter as a function of density. The increase in the kinetic energy of symmetric nuclear matter is also shown for comparison.} 
																	       \label{fig:kinetic}
																	       \end{center}
																	       \end{figure}
																	       A way of estimating the importance of correlations in a fermionic system is simply to evaluate the difference between the expectation value of the kinetic energy of the system and the energy of a
																	       free Fermi gas with the same density and constituents,
																	       \begin{equation}
																	       \Delta T= \langle T \rangle- E_{FFG}.
																	       \end{equation}
																	       The larger is the value of $\Delta T$ the more important is the role of the correlations. We show in Fig.\ \ref{fig:kinetic} the density dependence of $\Delta T$ for TP and NP neutron matter as well as for conventional symmetric nuclear matter (SM). The increase of $\Delta T$ in the three cases indicates, as expected, that correlations  become more and more important when the density of the system increases. Note that
																	       in the whole range of densities explored, $\Delta T_{SM} > \Delta T_{NP} > \Delta T_{TP}$, reflecting the fact that SM is always more correlated than neutron matter independently of its spin polarization state, and 
																	       that NP neutron matter is always more correlated than TP one. However, the effect of correlations on
																	       the kinetic energy of TP neutron matter can not be discarded. Note also that the difference $\Delta T_{SM} - \Delta T_{NP}$ is larger than the difference $\Delta T_{NP} - \Delta T_{TP}$ up to $\rho\sim 0.45$ 
																	       fm$^{-3}$. This can be interpreted as an indication that the spin dependence of the nucleon-nucleon correlations is less strong than its isospin one at least in the low and medium density region. 

																	       To get a more quantitative  idea of the spin dependence of the nucleon-nucleon correlations in the following we analyze the role played by the different terms of the nuclear force, and in particular the spin dependent ones,
																	       in the determination of $S^{\langle V \rangle}_{sym}$ and $L^{\langle V \rangle}_S$. To such end, we apply the Hellmann--Feynman theorem  to the separate contributions of the Av18 potential and the Urbana IX three-nucleon force. The Av18 has 18 components of the form $v_p(r_{ij})O^p_{ij}$ with
																	       \begin{widetext}
																	       \begin{eqnarray}
																	       O^{p=1,18}_{ij}&=&1, \,\, \vec \tau_i \cdot \vec \tau_j, \,\,  \vec \sigma_i \cdot \vec\sigma_j, \,\, 
																	       (\vec \sigma_i \cdot \vec \sigma_j )(\vec \tau_i \cdot \vec \tau_j), \,\, 
																	       S_{ij}, \,\, S_{ij}(\vec \tau_i \cdot \vec \tau_j), \,\, \vec L \cdot \vec S, \,\, 
																	       \vec L \cdot \vec S(\vec \tau_i \cdot \vec \tau_j), L^2, \nonumber \\ 
																	       &&L^2(\vec \tau_i \cdot \vec \tau_j), \,\, 
																	       L^2(\vec \sigma_i \cdot \vec \sigma_j), \,\,  L^2(\vec \sigma_i \cdot \vec \sigma_j)(\vec \tau_i \cdot \vec \tau_j), \,\, 
																	       (\vec L \cdot \vec S)^2, \,\, 
																	       (\vec L \cdot \vec S)^2(\vec \tau_i \cdot \vec \tau_j), \nonumber \\
																		       &&T_{ij}, \,\, (\vec \sigma_i \cdot \vec \sigma_j)T_{ij}, \,\, 
																	       S_{ij}T_{ij}, \,\, (\tau_{zi}+\tau_{z_j}) 
	\label{eq:av18}
	\end{eqnarray}
	\end{widetext}
	being $S_{ij}$ the usual tensor operator, $\vec L$ the relative orbital angular momentum, 
	$\vec S$ the total spin of the nucleon pair, and $T_{ij}=3\tau_{zi}\tau_{zj}-\tau_i \cdot \tau_j$ the isotensor operator defined
	analogously to $S_{ij}$. Note that the last four operators break the charge independence of the nuclear interaction.

	As we said above, the Urbana IX three-body force is reduced to an effective density-dependent two-body force when used in the BHF 
	approach. For simplicity, in the following we refer to it as reduced Urbana force. This force is made of $3$ components of the 
	type $u_p(r_{ij},\rho)O^p_{ij}$ where 
	\begin{equation}
	O^{p=1,3}_{ij}=1, \,\, (\vec\sigma_i\cdot\vec\sigma_j)(\vec\tau_i\cdot\vec\tau_j), \,\, S_{ij}(\vec\tau_i\cdot\vec\tau_j) \ ,
	\label{eq:uix}
	\end{equation}
	introducing additional central, $\sigma\tau$ and tensor terms (see {\it e.g.,} Ref.\ \cite{bf99} for details).

	The separate contributions of the various components of the Av18 potential and 
	the reduced Urbana force to the energy per particle of TP and NP neutron matter, and to $S^{\langle V \rangle}_{sym}$ and $L^{\langle V \rangle}_S$ at the empirical value of the nuclear saturation density are given in Tab.\ \ref{tab3}. Note that the largest contribution for both $S^{\langle V \rangle}_{sym}$ and $L^{\langle V \rangle}_S$ comes from the $\vec\sigma_i\cdot\vec\sigma_j$, $(\vec\sigma_i\cdot\vec\sigma_j)(\vec\tau_i\cdot\vec\tau_j)$ and $L^2$ terms.


	\begin{center}
	\begin{table}
	\begin{tabular}{crrrr}
	\hline
	\hline
	& $\langle V \rangle_{TP}$ & $\langle V \rangle_{NP}$ & $S^{\langle V \rangle}_{sym}$ & $L^{\langle V \rangle}_S$ \\
		\hline
		$\langle V_1 \rangle $    &  $-24.856$ & $-26.415$  & $1.559$   & $-3.012$    \\
			$\langle V_{\vec\tau_i\cdot\vec\tau_j} \rangle$     &  $-3.129$  & $-4.157$    & $1.028$  & $0.506$  \\
			$\langle V_{\vec\sigma_i\cdot\vec\sigma_j} \rangle$     &  $3.207$  & $-0.438$   & $3.645$   & $9.147$    \\
			$\langle V_{(\vec\sigma_i\cdot\vec\sigma_j)(\vec\tau_i\cdot\vec\tau_j)} \rangle$     &  $13.046$  & $-5.470$  & $18.516$   & $50.328$    \\
			$\langle V_{S_{ij}} \rangle$     &  $-0.980$  & $-0.608$    & $-0.372$  & $-1.075$   \\
			$\langle V_{S_{ij}(\vec\tau_i\cdot\vec\tau_j)} \rangle$     &  $-5.725$  & $-4.219$  & $-1.506$  & $-3.625$   \\
			$\langle V_{\vec L\cdot \vec S} \rangle$     &  $-8.638$  & $-6.076$   & $-2.562$  & $-2.855$  \\
			$\langle V_{\vec L\cdot \vec S (\vec\tau_i\cdot\vec\tau_j)} \rangle$     &  $-3.090$  & $-2.148$   & $-0.942$   & $-3.303$    \\
			$\langle V_{L^2} \rangle$     &  $14.090$  & $9.188$   & $4.902$  & $18.735$    \\
			$\langle V_{L^2(\vec\tau_i\cdot\vec\tau_j)} \rangle$  &  $-2.899$  & $-2.142$    & $-0.757$  & $-3.238$  \\
			$\langle V_{L^2(\vec\sigma_i\cdot\vec\sigma_j)} \rangle$  &  $1.410$   & $1.016$    & $0.394$  & $0.741$    \\
			$\langle V_{L^2(\vec\sigma_i\cdot\vec\sigma_j)(\vec\tau_i\cdot\vec\tau_j)} \rangle$  &  $-0.787$   & $0.017$    & $-0.804$   & $-5.024$   \\
			$\langle V_{(\vec L\cdot \vec S)^2} \rangle$  &  $5.652$   & $3.262$    & $2.390$  & $12.803$   \\
			$\langle V_{(\vec L\cdot \vec S)^2(\vec\tau_i\cdot\vec\tau_j)} \rangle$  &  $6.903$   & $4.032$   & $2.871$  & $14.275$    \\
			$\langle V_{T_{ij}} \rangle$  &  $0.006$   & $0.002$    & $0.004$  & $0.022$   \\
			$\langle V_{(\vec\sigma_i\cdot\vec\sigma_j)T_{ij}} \rangle$  &  $-0.013 $  & $-0.015$   &  $0.002$  & $-0.010$    \\
			$\langle V_{S_{ij}T_{ij}} \rangle$  &  $0.004$   & $0.003$    &  $0.001$ & $-0.102$   \\
			$\langle V_{(\tau_{z_i}+\tau_{z_j})} \rangle$  &  $-0.055$   &  $-0.070 $   &  $0.015$ & $-0.054$   \\
			\\
			$\langle U_1 \rangle$   &  $-0.019$   &  $1.744$   &  $-1.763$ & $-6.967$  \\
			$\langle U_{(\vec\sigma_i\cdot\vec\sigma_j)(\vec\tau_i\cdot\vec\tau_j)} \rangle$  &  $-0.922$   &  $-0.708$   &  $-0.214$ & $-0.872$   \\
			$\langle U_{S_{ij}(\vec\tau_i\cdot\vec\tau_j)} \rangle$  &  $2.011$  &  $2.152$  &  $-0.141$  & $-0.506$   \\ 
			\hline
			\hline
			\end{tabular}
			\caption{Contributions of the various components of the Av18 potential (denoted as $\langle V_i \rangle$) and the reduced Urbana force (denoted as $\langle U_i \rangle$) to the total energy per particle of TP and NP neutron matter and to the spin symmetry energy and its slope parameter at the empirical saturation density of symmetric nuclear matter $\rho_0=0.16$ fm$^{-3}$. Results are given in MeV.} 
			\label{tab3}
			\end{table}
			\end{center}

			As we have already seen in Table II, the total interaction energy for TP neutron matter  is in absolute
			value much smaller than for the NP one. This is the result of strong cancellations between the contributions of the different pieces of the potential. The contributions to $S_{sym}^{<V>}$ and $L_S^{<V>}$ 
			are important when there is a difference in the behavior of the interaction betwen TP and NP neutron matter. For instance, the contribution of the central part $\langle V_1\rangle$ is very similar in TP and NP neutron matter and therefore its contribution to $S_{sym}^{<V>}$ is small.  Relevant contributions are associated to 
$\langle V_{\vec\sigma_i\cdot\vec\sigma_j} \rangle$,
$\langle V_{(\vec\sigma_i\cdot\vec\sigma_j)(\vec\tau_i\cdot\vec\tau_j)} \rangle$ and also 
			to $\langle V_{L^2} \rangle$. On the other hand the contributions of the three-body forces to the 
			spin symmetry energy are moderatly small and of negative sign, at $\rho_0$. 


			\section{Summary and conclusions}

			We have calculated the kinetic and potential energy contributions of the spin symmetry energy of neutron matter using the realistic Argonne Av18 two-body interaction  suplemented with the   Urbana IX three-body force averaged to provide a two-body density dependent one suitable to be used in BHF calculations. It has been shown that this realistic interaction do not favour a ferromagnetic transition of neutron matter. As the symmetry energy, the spin symmetry energy is an increasing function of density, at least in the range of densities considered. Both, the kinetic and the potential energy contributions, {\it i.e.,} the difference of these energies between polarized and normal neutron matter, are positive in the full range of densities considered. 

			The contributions of the different pieces of the interaction and 
 its partial wave decomposition allows to understand the origin of the different role
 of the interaction in TP and NP neutron matter. In most of the cases, the Pauli principle, which forbids the interaction in certain partial waves in  totally polarized neutron matter is the origin of most of the differences. The main contribution comes from the S=0 forbidden channels in TP neutron matter, in particular from the $^1$S$_0$ and $^1$D$_2$ partial waves.  On the other hand, three-body forces play a secondary role in the determination of the spin symmetry energy. 

Finally, we  have quantitatively established that NP  neutron matter is more correlated than TP one by looking at the difference of their kinetic energies and the corresponding ones of their underlying Fermi seas. In spite of being less correlated, however, the role of correlations in totally polarized neutron matter cannot be ignored when using realistic interactions. 

\section*{Acknowledgements}

This work is supported by Grant No. FIS2014-54672-P from MICINN (Spain), and Grant No. 2014SGR-401 from Generalitat de Catalunya (Spain), the GSI-TU Darmstadt bilateral
cooperation agreement and by the  Helmholtz International Center for FAIR, and by NewCompstar, COST Action MP1304.

	\end{document}